\documentclass[prl,aps,epsfig,twocolumn,showpacs]{revtex4}
\usepackage{epsfig,amssymb}

\newcommand\beq{\begin{equation}}
\newcommand\eeq{\end{equation}}
\newcommand\bea{\begin{eqnarray}}
\newcommand\eea{\end{eqnarray}}
\newcommand\non{\nonumber}

\begin{document}

\textheight=23.8cm

\title{\Large Adiabatic charge pumping through a dot at the junction of
$N$ quantum wires}
\author{\bf Shamik Banerjee, Anamitra Mukherjee, Sumathi Rao and Arijit Saha} 
\affiliation{
\it Harish-Chandra Research Institute, Chhatnag Road, Jhusi, Allahabad 
211019, India }
\date{\today}

\begin{abstract}

We study adiabatic charge pumping through a quantum dot placed at the
junction of $N$ quantum wires. We explicitly map out the pattern of 
pumped charge as a function of the time-varying tunneling parameters
coupling the wires to the dot and the phase between any two 
time varying parameters controlling the
shape of the dot. We find that with $N-2$ time-independent well-coupled leads,
the maximum pumped charge in the remaining  two leads is strongly
suppressed with increasing $N$, leading to the possibility of
tuning of the pumped charge, by modulating the coupling of the $N-2$ leads.

\end{abstract}
\maketitle

An adiabatic charge pump\cite{THOULESS} was theoretically proposed by Thouless
for an isolated system and later extended to open systems at finite
temperature\cite{ALTGLAZ,BPT,BROUWER}. It is essentially a device that 
drives current through a system at zero bias 
by a time-dependent  variation of two or more parameters.
Although theoretically proposed about  a decade ago, 
Switkes {\it et
al}\cite{SWITKES} recently
demonstrated that  an adiabatic electron pump,
which produces a dc voltage in response to a cyclic change in the 
confining potential of an {\it open} quantum dot was experimentally feasible. 
This has led to renewed 
interest\cite{KOUW,AVR,LEV,B,AB,AS,AVRON,ALEINER,EWAA1,EWAA2,SHARMA} 
in the field.

In this paper, we study adiabatic charge pumping through a quantum dot
placed at a junction of $N$ quantum wires and which is well-coupled
to the  wires. We show that under a time dependent variation of 
two of the
tunneling parameters to the dot, charge $Q_{ij} =
Q(i,\tau)-Q(j,\tau)$ can be
pumped from wire $i$ to wire $j$. 
An important point to note is that unlike the case when $N=2$,
the transmission between any two wires 
for $N \ge 3$ never reaches unity anywhere in the parameter space, 
because of the possibility of leaking out through the other wires.
Thus, analogously to the $N=2$ case\cite{EWAA1},
large pumping occurs when the 
 `pumping contour' (a closed curve in the $J_i$-$J_j$ plane, 
where $J_i$ and $J_j$ denote the tunneling parameters between wire $i,j$ 
and the dot) encloses transmission maxima peaks (when $\sum_{j}T_{ij}=1$)
and cuts the resonance line at points where the
transmission is small, thus ensuring that most of the resonance
line is enclosed.

Consider a
time-dependent scatterer connected to $N$ single channel  leads. 
In the adiabatic limit, at leading order, ($O(\omega^0)$, where $\omega$ is
the frequency),
the incoming particle sees a static
scatterer, and hence the scattering and the current 
can be computed using time-independent scattering theory
(the Landauer-Buttiker formula). But as shown by Buttiker, Pretre and
Thomas\cite{BPT} and Brouwer\cite{BROUWER},
even at order $O(\omega^1)$, the current and the charge pumped can
be computed from the `frozen' or `quasi-static' 
scattering matrix\cite{AVRON} data.
For a simultaneous slow and small variation of two 
time-dependent parameters
$X_1$ and $X_2$, the charge emitted from wire $i$ is given by 
\bea
Q(i,\tau) 
&=& {e\over 2\pi} \int_0^\tau [
|r_{ii}|^2 \frac {d\theta_i}{dt} + \sum_{j\ne i}^N |t_{ij}|^2 
\frac {d\psi_{ij}}{dt}] \\
&=&{e\over 2\pi} \int_0^\tau \frac {d\theta_i}{dt} - 
\sum_{j=1}^N |t_{ij}|^2  (\frac {d\theta_i}{dt}~ -
\frac {d\psi_{ij}}{dt}~)
\label{pcharge3}  
\eea
where the first term is denoted as $Q_{int}$ and the second as $Q_{tran}$
and 
the $S$-matrix for an $N$-lead system is given by
\beq
S = \left(\begin{array}{llll}
|r_{11}|e^{i\theta_1} & |t_{12}|e^{i\psi_{12}} & |t_{13}|
e^{i\psi_{13}} & \dots \\
|t_{21}|e^{i\psi_{21}} & |r_{22}|e^{i\theta_2} & |t_{23}|e^{i\psi_{23}} 
&\dots \\
|t_{31}|e^{i\psi_{31}} & |t_{32}|e^{i\psi_{32}} & |r_{33}|e^{i\theta_3} 
&\dots \\
\dots & \dots & \dots & \dots 
\end{array} \right)~.
\eeq
$r_{ii},t_{ij}$, $\theta_i$ and $\psi_{ij}$
are time dependent in general.
We will mainly use Eq.(\ref{pcharge3}) in our computations. 
Note that the first term on the RHS of Eq.(\ref{pcharge3}) is a total
derivative and denotes the winding number around the $R=0$ singularity.
Thus $Q_{int}$ 
is forced to be an integer. The second term is not quantised, however, and
hence, the pumped charge is not an integer.  But by choosing the contour
appropriately, to cut the transmission resonance maxima when it is
small, $Q_{tran}$ can be made small, and the charge pumped is
almost quantised\cite{EWAA1}. As we shall see, this is harder to arrange
when $N>2$, because there is a possibility of transmission into
the other wires. This explains why the value of the maximum pumped
charge decreases as $N$ increases, which is one of the main results
of this report.

Let us now write down the lattice model for a quantum dot at the
junction of $N$ quantum wires.  The wires are modeled as a one-dimensional
chain of sites, with on-site energy zero and with nearest neighbour  hopping
amplitudes given by $-J$. The dot is again modeled as a bunch of tight-
binding sites with nearest neighbour hoppings given by $-J_D$ and on-site
energy $\epsilon_0$. Finally, the dot is coupled to the $i^{th}$ lead
by a hopping element $J_i$. In the work presented in this brief
report, we have taken $J_D=J$ and $\epsilon=0$ for simplicity,
since taking $J_D\ne J$ and $\epsilon \ne 0$ makes no qualitative
difference to the results\cite{us}.
Thus, the Hamiltonian of the system is given
by 
\bea
H &=& H_{\rm wires} + H_{\rm dot-wire} + H_{\rm dot} \non \\
&=&  \sum_{i=1}^N \big[-J
\sum_{m=l}^L c_{im}^\dagger c_{im+1} -J_i c_{il}^\dagger c_{il+1}  \non \\
&& -J \sum_{m=1}^{l-1} c_{im}^\dagger c_{im+1} -J c_0^\dagger c_{i1} +h.c.
\big]
\eea
Here $L$ denotes the number of sites on each wire with wire
index $i$ and $l$ denotes the
number of sites on each wire representing the dot. $c_0$ denotes
the dot point which is common to all the wires. 
Since our focus here is on charge pumping, we have not included spin
in the above Hamiltonian. 

For fixed values of $J$, and $J_i$, ($i=1,\dots N$),
the scattering problem is
straightforward. Let us assume that the incoming wave is along wire $1$,
without loss of generality. Then
the wave-function at the site $n$  in lead $1$ is given by
$
c_{in}=\exp(-ikx_{n})+r_{11}\exp(ikx_{n})  ;  (~n\ge l+1)
$
and the wave-function at site $n$ in all other leads $i=2,\dots,N$ are given by
$
c_{in}=t_{1i}\exp(ikx_{n})  ; (~ n \ge l+1)  
$
where $x_n=na$ for the lattice spacing $a$.
Note that the numbering of the sites, implies that incoming
waves are denoted by $\exp(-ikx_{n})$ whereas outgoing states
are denoted by $\exp(ikx_{n})$. The wave-function at sites within the dot
is given by ($j=1,\dots N$)
$
a_{jn}=P^j_{1}\exp(-iqx_{n})+P^j_{2}\exp(iqx_{n}) ; (~n\le l).
$
The wave-vectors $k$ and $q$ are related to the energy as
\beq
E_k=-2J\cos ka \quad {\rm and} \quad E_k = \epsilon_0-2J_D\cos qa
\label{doteq2}
\eeq 
as can be seen by solving the equations in the bulk of the wire and
the dot.
The set of boundary equations (for amplitudes at the boundary of the 
dot and the wire, $i.e.$, sites $l$ and $l+1$ 
and at the junction point of the dot) 
can be solved to obtain $r_{11}$
and $t_{1i}$ in terms of $J,J_D$ and $J_i$, 
but it gets progressively more difficult for higher values of $N$
to get analytical results. 
However, the equations  can be solved numerically and we can obtain the 
scattering $S$-matrix in the parameter space of the hopping amplitudes.

To obtain pumped charge, any two of the hopping parameters that couple the dot
to the wire need to be adiabatically varied.  Hence, in the lattice
model, two of the hopping parameters that couple the dot to the
wires are taken to oscillate with the frequency $\omega$, with a
modulation parameter $P_i$ and a phase difference $2\phi$, $i.e.$, we
choose
\bea
J_i &=& J_{i0} + P_i \cos(\omega t +\phi) \nonumber\\
J_{j\ne i} &=& J_{j0} + P_j \cos(\omega t-\phi)~,
\label{jijj}
\eea
where $j$ is one value not equal to $i$. In this report, we choose
$P_i=P_j$. The case where $P_i\ne P_j$, as also the cases when more
than two parameters are varied will be considered elsewhere\cite{us}.

The charge pumped out from wire $i$, $Q(i,\tau)$ 
is then computed from the time dependence of
the $S$-matrix parameters $r_{ii}$, which, in turn, are
known in terms of the hopping parameters,
using Eq.\ref{pcharge3}.  Finally, we can obtain the charge
pumped between different wires as
\beq
Q_{ij} = Q(i,\tau) - Q(j,\tau)~.
\eeq
Note that $Q(i,\tau)$ is computed with the wave incident in wire $i$,
whereas $Q(j,\tau)$ is computed with the wave incident in wire $j$.
The results for the 
pumped charges for the different cases are given below.  

\begin{figure*}
\begin{center}
\epsfig{figure=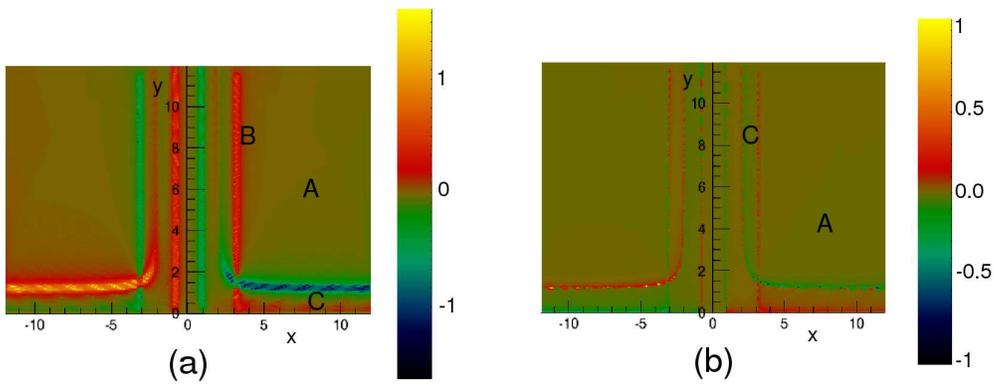,width=15.0cm}
\end{center}
\caption{Contour plot of the pumped charge for $N=2$ and $3$ wires
going from left to right.
The $x$- and $y$- axes are $P\cos\phi$ and $P\sin\phi$. The colour scale
is on the right of each plot. Note that the pattern remains the same
in both ; for $N=2$, the pumped charge reaches a maximum of 
$\sim \pm 2$, whereas for $N=3$, it reaches a maximum of $\pm 1.11$. 
}
\label{dotqpfig1}
\end{figure*}

\begin{figure}
\begin{center}
\epsfxsize 3.0 in
\epsfysize 1.7 in
\epsfbox{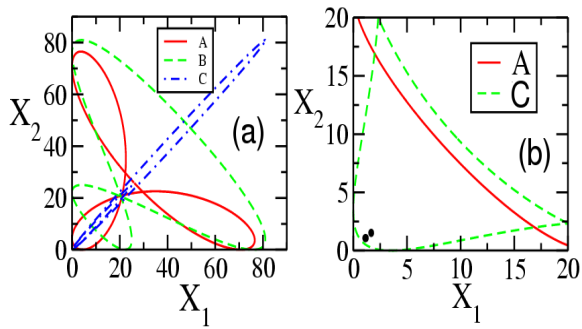}
\end{center}
\caption{Typical pumping contours 
 as a function of the pumping parameters 
$X_1=J_1^{2}$ and $X_2=J_2^{2}$. 
For $N=2$, the full pumping contours are  shown for three classes 
corresponding to the regions $A$,$B$ and $C$ in Fig. 1(a).
For $N=3$, only a small
part of contours of classes $A$ and $C$ corresponding to regions
$A$ and $C$ in Fig. 1(b) is 
shown, along with the resonance points $R=0$. Note that  contour $A$ 
does not enclose the resonance points  and contour $C$ encloses both once}
\label{dotqpfig2}
\end{figure}

 The parameters that affect the value
of the pumped charge are 
the pumping amplitude $P$ and the phase $\phi$. We have
varied $\phi$ over its full range.  Within the adiabatic approximation,
the pumped charge is independent of $\omega$ and so we have chosen
a specific value of $\omega$. 

\begin{figure}[h]
\begin{center}
\epsfxsize 3.0 in
\epsfysize 1.7 in
\epsfbox{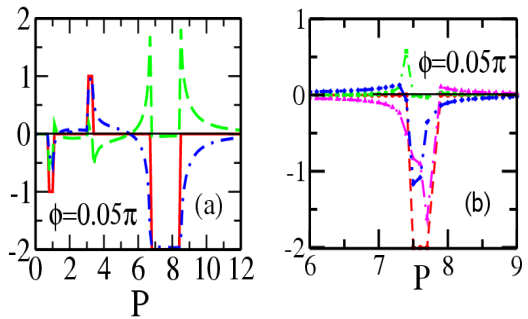}
\end{center}
\caption{The different contributions to the pumped charge $Q$ for 
$N=2$ (a) and $N=3$ (b).  The full red line is $Q_{int}$, 
the blue dash-dot line is the total pumped charge $Q$
and the green (and magenta for $N=3$) dotted lines are $Q_{tran}$. Note
that large pumped
charge is correlated with $Q_{int}\ne 0$. In Fig.3(b), only a small region
of $P$ is shown, because the width of $P$ for which pumped charge is large
is smaller.The numerical values chosen to obtain $Q$ are 
$J_{D}=J=1.0,J_{10}= J_{20} =2.0,\phi=0.05\pi
 ka=0.001 \pi,
\omega=1.0$.  }
\label{dotqpfig3}
\end{figure}

\vskip 0.0cm


We have solved the set of boundary equations for $N=2,3$ and $4$ wires.
For $N=2$, we have chosen $lN=4$, and both
the hopping amplitudes coupling the dot to the wire are made time-dependent 
using Eq.\ref{jijj}, and we obtain a single pumped charge $Q=
\Delta Q_{12} -\Delta Q_{21}$. For 
$N=3$, any two of the three hopping amplitudes coupling
the dot (with $lN+1=4$ sites) to the wire can be made time-dependent.
Let us call them $J_1$ and $J_2$. 
We then obtain the pumped charge $Q 
=\Delta Q_{12} -\Delta Q_{21}$. (Similarly, by making the other
hopping amplitudes time dependent, we can obtain 
$Q'= \Delta Q_{23} -\Delta
Q_{32}$ and $Q'' =\Delta Q_{31} -\Delta Q_{13}$).  
The pumped charge $Q$ is computed 
when the third hopping amplitude $J_3=1.1$
which is of the same order  as the mean values of $J_{1,2}$ 
within each cycle.

We have plotted the pumped charge $Q$ as a function
of the pumping strength $P_{1,2}=P$ and the phase
$\phi$ 
in the contour plots of Fig.\ref{dotqpfig1} for $N=2$ and 3.
We have chosen $\phi$ to go from $0 \rightarrow \pi$, since
Eq.\ref{jijj} shows that the  difference in the phase between
the tunneling parameters on the two leads connected to the dot, 
is given by $2\phi$, and we have chosen a maximum of 10 for $P$.
In Fig.\ref{dotqpfig2}, we have plotted 
the different classes of contours in the $P-\phi$ plane for $N=2$
and 3.
We find that 
the contours 
can be classified into 3 catagories.  Those which (A) do 
not enclose either  of the two $\vert T \vert=1 $ points, 
(B) includes one of them more number of times than the other and 
finally (not shown for $N=3$) 
(C) includes both of them equal number of times.  
We see (as was shown in Ref. \cite{EWAA1}) that the pumped charge is 
large only 
in the case (B) with the sign of the pumped charge depending on the sense 
in which the contour encloses the $ \vert T \vert=1 $ point.
We note that the pattern of pumped charge remains the
same for $N=2$ and 3, since the form of the variation of the
paramters remains the same. But the magnitude of the pumped charge decreases
for $N=3$. This is due to the 'leak' of 
the wavefunction in the third  direction, so that 
the transmission never reaches unity in the 1-2 direction.

In Fig. 3, we have separately shown  the contributions of the 
the winding number term $Q_{int}$ and the 
transmission term $Q_{tran}$(see Eq.(\ref{pcharge3}) for $N=2$ and 3.
These  graphs clearly show that large pumped
charge is always correlated with the winding number term being non-zero.
Fig. 3(b)  also shows that the range of values of $P$ for which the pumped
charge is large is narrower for $N=3$. This can be correlated
with the coming together of the resonance lines  as the coupling
to the third wire is increased. This is shown explicitly in Fig. 4(a),
and also in Fig.2(b), where it is clear that it is hard to find
contours which enclose one resonance but not the other.
Thus, the parameter range ($P_c$, $\phi$),
for which the contours enclose one maxima and not the other is quite
small, and so the region where pumped charge is large is small. 

In Fig.\ref{dotqpfig4}(b), we have shown the
plots of the pumped charge when $J_3\ll J_{1,2}$ and $J_3\gg J_{1,2}$. 
In either case, it can be checked
that the plots are very similar to the $N=2$ case studied
earlier \cite{EWAA1}.  This
is not surprising because both for $J_3$ very small and
$J_3$ very large, the leak to the third wire is very small.
But for intermediate $J_3$, we see that not only is the maximum pumped
charge smaller, it occurs for a narrower range of parameters as expected.

\vskip 0.0cm
\begin{figure}[h]
\begin{center}
\epsfxsize 2.2 in
\epsfysize 1.7 in
\epsfbox{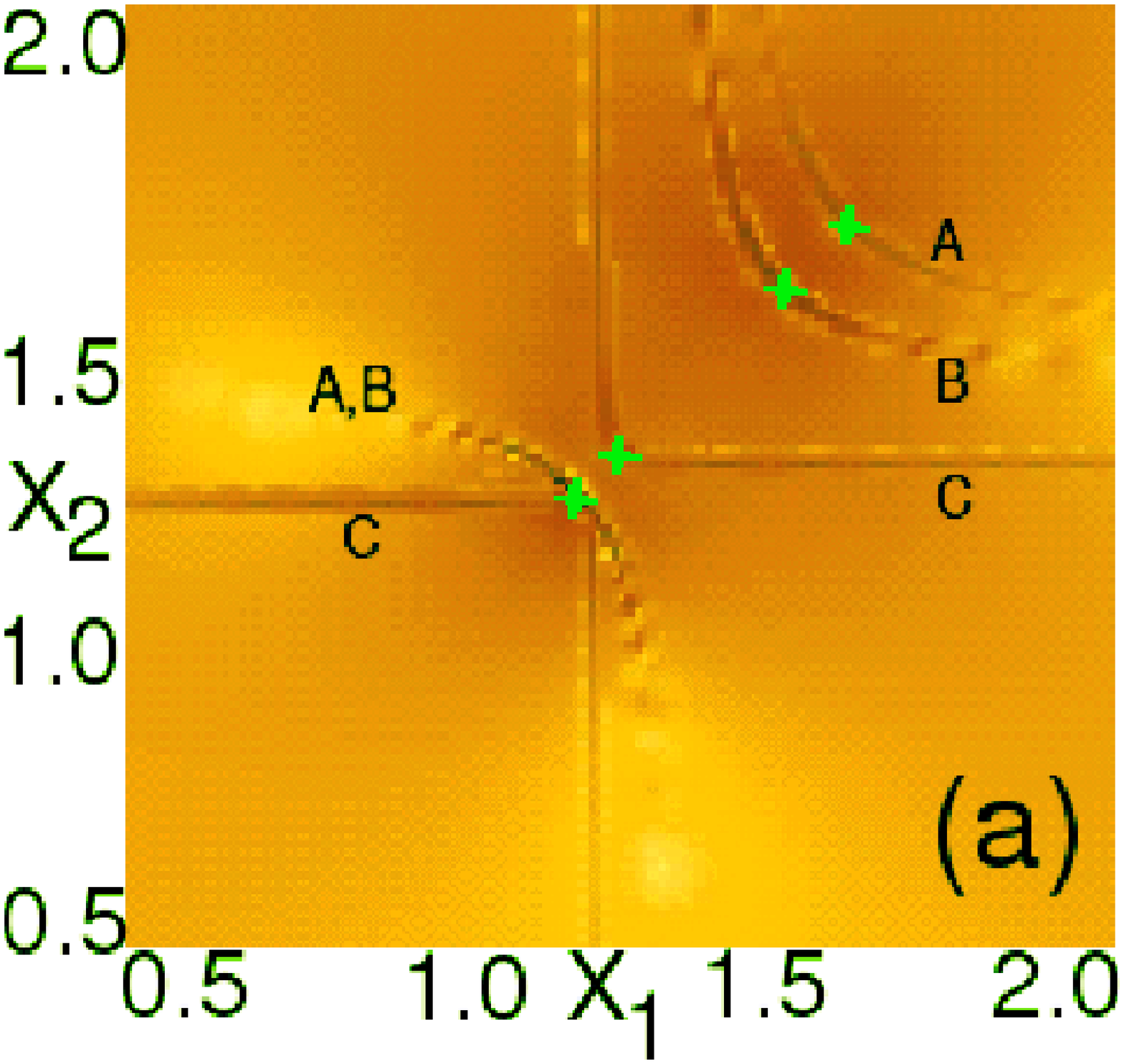}
\end{center}
\end{figure}
\begin{figure}
\begin{center}
\epsfxsize 2.2 in
\epsfysize 1.7 in
\epsfbox{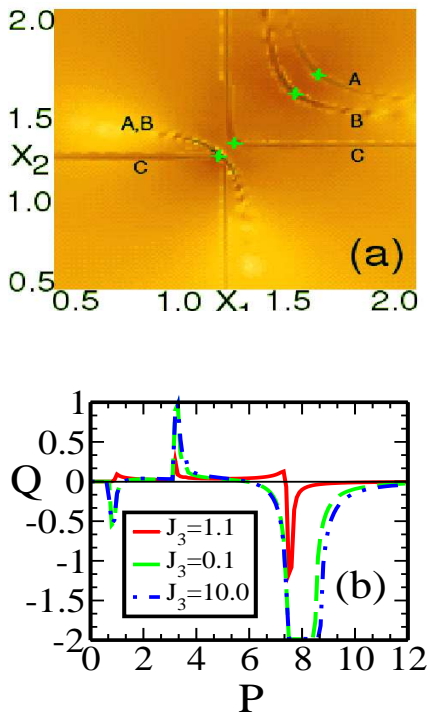}
\end{center}
\caption{(a) Contour plot
of transmission maxima $ (T_{12}+T_{13})$ for three values of $J_3$. 
Lines $A$, $B$ and $C$ denote resonance lines for $J_3=0.1. 0.9$
and $1.1$ respectively. The crosses denote the $R=0$ points. Note that
in the lower left quandrant, lines $A$ and $B$ are indistinguishable
and the resonance point on $A$,$B$ and $C$ is indistinguishable and
denoted by a single cross.
(b) Pumped charge $Q$ as a function of the pumping amplitude
$P$ for three values of
$J_3 =0.1,1.1$ and $J_3=10.0$.
The other parameters are the same as in Fig.\ref{dotqpfig2}.}
\label{dotqpfig4}
\end{figure}

\begin{figure}[h]
\begin{center}
\epsfxsize 2.0 in
\epsfysize 1.25 in
\epsfbox{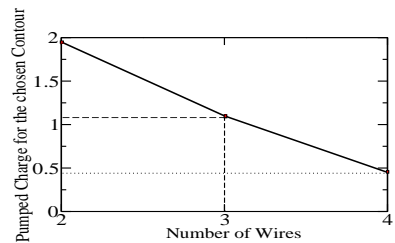}

\end{center}
\vspace{0.15cm}
\caption{Maximum pumped charge versus the number of wires. }
\label{dotqpfig5}
\end{figure}

\vskip 0.0cm
What happens when 4 wires are
attached to the dot and two of the hopping amplitudes
coupling the wire to the dot are made time-dependent?
Can we  identify trends as
the number of wires $N$ increases? To check this, we 
obtained the numerical results for $N=4$ wires
and we find that the magnitude of pumped charge
decreases, with regions of large pumped charge getting
thinner as we increase $N$. 
In Fig.\ref{dotqpfig5}, we plot the maximum pumped charge  as a function
of the number of wires well-coupled (and held constant in time)
to the dot.

Hence, we conclude that it is harder to tune for  `almost
quantised' pumped charge for pumping through a dot at a junction 
of $N$ wires except 
for $N=2$.  However, the magnitude of pumped charge is not vanishingly small
for $N = 3,4$. There is a suppression of the maximum 
pumped charge from $\sim 2$  for $N = 2$ to about 1.2  for $N=3$  and to 0.45
 for $N=4$. 
This can be easily explained as a `leak' of the electrons into the wires
that are not participating in the pumping process. Hence, it is
easy to see that no pumped charge is expected in the large $N$ limit.


\vspace{0.15cm}

In conclusion, we have explicitly obtained the pumped charge 
as a function
of the various parameters such as pumping amplitude, phase $\phi$,
pumping frequency, etc, for pumping 
through a quantum dot placed at the junction of $N$ wires where
$N=2,3$ and 4.  We note that the 
maximum pumped charge reduces in magnitude as the number of wires is 
increased because some of the charge can leak out through
the other wires which are  not involved in the pumping process. 
The pumped charge is also seen to be large only when the pumping contour
encloses the transmission maxima (not necessarily a $\vert T \vert=1$ point)
appropriately and cuts the resonance line at a point
where the transmission is low. 

More general cases which are possible for $N\ge 3$, which involve
time dependent variation of all the tunneling parameters coupling
the wire to the dot can also be studied\cite{us}.
One might expect that unless the time-dependences are carefully adjusted
the pumped charge would be quite low. However, they have 
not been explicitly considered here. 
The extensions to higher $\omega$, where the adiabatic
approximation breaks down will also be presented elsewhere. 

\vskip 0.5cm

\centerline{\bf ACKNOWLEDGEMENTS}

\acknowledgments

We acknowledge use of the Beowulf cluster at the Harish-chandra
research Institute in our computations.

\end{document}